\documentclass[11pt,a4paper]{article}
\usepackage[utf8]{inputenc}
\usepackage{amsmath}
\usepackage{amssymb}
\usepackage{amsfonts}
\usepackage[citecolor=blue,colorlinks=true,linkcolor=blue]{hyperref}
\usepackage{caption}
\usepackage{subcaption}
\usepackage{graphicx}
\usepackage{geometry}
\usepackage{hyphenat}
\usepackage{csquotes}
\usepackage[english]{babel}

\defineshorthand{~=}{\hyp{}}
\usepackage{bm}
\usepackage[parfill]{parskip}
\newcommand{\ts}{\textsuperscript}
\usepackage[noblocks]{authblk}

\title{\sc Calculation of Epidemic First Passage and Peak Time Probability Distributions}

\author[1,2*]{Jacob Curran~=Sebastian}
\author[2,3]{Lorenzo Pellis}
\author[2,3]{Ian Hall}
\author[2,3]{Thomas House}
\affil[1]{Section of Epidemiology, Department of Public Health, University of Copenhagen, Copenhagen, Denmark}
\affil[2]{Department of Mathematics, University of Manchester, Manchester, UK}
\affil[3]{Alan Turing Institute for Data Science and Artificial Intelligence, London, UK}
\affil[*]{Corresponding Author: jacob.curran-sebastian@sund.ku.dk}

\date{}

\begin{document}

\maketitle

\begin{abstract}
\noindent{}Understanding the timing of the peak of a disease outbreak forms an important part of epidemic forecasting. In many cases, such information is essential for planning increased hospital bed demand and for designing of public health interventions. The time taken for an outbreak to become large is inherently stochastic, and therefore uncertain, but after a sufficient number of infections has been reached the subsequent dynamics can be modelled accurately using ordinary differential equations. Here, we present analytical and numerical methods for approximating the time at which a stochastic model of a disease outbreak reaches a large number of cases and for quantifying the uncertainty arising from demographic stochasticity around that time. We then project this uncertainty forwards in time using an ordinary differential equation model in order to obtain a distribution for the peak timing of the epidemic that agrees closely with large simulations but that, for error tolerances relevant to most realistic applications, requires a fraction of the computational cost of full Monte Carlo approaches. 
\end{abstract}

\section{Introduction}
\label{Intro}
The COVID-19 pandemic, which began in late 2019 with an outbreak of the novel SARS-CoV-2 pathogen in Wuhan, China, has underscored the need for mathematical modelling that can quickly and accurately estimate important epidemiological quantities and provide meaningful forecasts of disease dynamics. Insights from modelling are particularly crucial for planning responses and interventions in the early stages of an epidemic \cite{overton2020using}. Producing epidemiological forecasts and estimating key parameters, such as the generation time, doubling time and the basic reproduction number, $R_0$, while an epidemic is ongoing is subject to enormous uncertainty \cite{pellis2021challenges, kucharski2020early, silk2022uncertainty}. Stochastic models of disease transmission offer an advantage in that they are able to capture the randomness of events that occur during an outbreak, particularly when that outbreak is in its early stages \cite{you2020estimation, abbott2020transmissibility}. Not accounting for such stochasticity can lead to discrepancies between model outputs and outbreak data. Furthermore, if super-spreading events and extinction of individual transmission chains are common in outbreaks of a given disease, then models that do not capture the random nature of these events can create biased estimates of key epidemiological quantities \cite{lloyd2005superspreading}. Fitting deterministic models to disease data fails to fully capture the underlying uncertainty in the disease dynamics, effectively attributing any discrepancy between the model and data to measurement error \cite{read2021novel}. Relying on deterministic models alone can therefore lead to bias in the estimation of epidemiological parameters, as well as an underestimation of the true uncertainty in model outcomes \cite{king2015avoidable} and so stochastic models are preferred wherever possible \cite{andersson2012stochastic}.

Stochastic models also have an advantage in modelling the early phases of epidemics in that they allow events to occur after a random time, thus accounting for the large variability in the time taken for an outbreak to begin growing exponentially. Hybrid modelling that incorporates both stochastic and deterministic elements have been used previously to enrich deterministic models with uncertainty in both the time to extinction and probability of extinction of an outbreak \cite{yan2018distribution, binder2012hybrid, rebuli2017hybrid} as well as the total duration of an epidemic \cite{barbour1975duration}. In the context of COVID-19, stochastic models have been used to generate a distribution of starting times, together with initial conditions, for a deterministic model that only begins once an epidemic has reached its exponential growth phase \cite{dyson2021possible}. There have also been approaches to prediction under uncertainty of the peak timing and height from incidence data collected in the early phase of an outbreak with parametric regression models that have been used during the COVID-19 pandemic \cite{alaimo2021nowcasting}.

However, simulating sample trajectories from a stochastic model is computationally expensive, particularly when the number of events is large, and also has the disadvantage that it does not offer any mathematical insight into the true underlying distribution of the number of cases at a given time. This makes large stochastic simulations particularly unsuitable for model calibration or sensitivity analysis, for which many such simulations may be needed in order to test different regions of a given parameter space. This is particularly relevant when many potential outbreak scenarios need to be considered, as was the case for modelling the roadmap for lifting mass lockdown restrictions in the United Kingdom \cite{keeling2022comparison}. The tools outlined in this paper are intended to enrich deterministic modelling of epidemics with explicit consideration of stochastic effects in the early growth phase, in a way that provides more tractable insight and is computationally much more efficient than running large outbreak simulations.  

We work with a simple stochastic version of the Susceptible-Infectious-Recovered (SIR) model, using a continuous-time single-type branching process to approximate early behaviour, and use insights from this model to define the time after the initial case is infected at which the dynamics of the subsequent epidemic are well-approximated by a deterministic model. Intuitively, stochasticity becomes negligible when the number of infectives is sufficiently large, so we aim to describe the distribution of the time, $T$, at which the population, $Z(t)$, of the branching process crosses a particular size $Z^*$. To the best of our knowledge, no analytical results for this First Passage Time (FPT) are currently known. We identify a suitable $Z^*$ by imposing conditions on the population distribution at a given time, $T^*$ at which we argue stochasticity has become negligible due to both the probability of having zero cases being close to zero (provided that the outbreak has not already become extinct), and also that the standard deviation in the number of cases grows proportionally with the mean. We formalise these conditions in Section \ref{Tstardef}. We then define $Z^*$ as the mean population of the branching process at time $T^*$.

Given the lack of results on FPT distributions for branching processes, we consider a diffusion approximation to the branching process studied by Feller \cite{feller1951genetics, feller1951} for which the FPT distribution can be investigated. The Feller diffusion has a known distribution at each time point, and so we use existing results to obtain the FPT distribution for the number of cases in the Feller approximation to reach the same level as that of the branching process at time $T^*$. Finally, we consider an alternative approximation for the FPT distribution of the Feller diffusion using a Gaussian Process, which has the advantage of ease of implementation and applicability to a wider range of contexts than the epidemic models considered in this paper and discuss the benefits and limitations of each approach.

\section{Methods}
\subsection{Stochastic Model of Early Growth} \label{BP_outbreak}
We consider an SIR model of an outbreak of an infectious disease in a homogeneously mixing population that is fully susceptible to infection. We denote by $\beta$ and $\gamma$ the constant infection and recovery rates, respectively, and linearise the early epidemic growth phase, i.e.\ we ignore the depletion of the susceptible population. We do not currently consider any immigration of cases from an external source, so the population is assumed to be closed, and we do not consider transmission or recovery rates that depend on time, though this is the subject of ongoing and future work.

Based on these assumptions, our aim is to define the number of cases $Z^*$ in an outbreak which is large enough that the subsequent dynamics can be well-approximated by a deterministic SIR model. We then wish to quantify the uncertainty around this time by estimating the first-passage time distribution to $Z^*$. To define $Z^*$ we identify a time $T^*$ that is sufficiently late for a supercritical branching process to be ``far enough’’ from zero, and then define $Z^* = Z(T^*)$, where $Z(t)$ denotes the number of active infectious cases at time $t$ in our branching process.

We model the linearised early epidemic growth phase with a birth and death process, where these two events correspond to infection and recovery of infected individuals and hence occur at rates $\beta$ and $\gamma$, respectively.
This birth and death process can be equivalently described as a branching process in which individuals die at the time when either event occurs (the lifetime is exponentially distributed with rate $\beta+\gamma$) and is either replaced by two new individuals in the former event and zero individuals in the latter \cite{dorman2004garden}.
Therefore, at each event in the branching process model, an infectious case produces a number of offspring that is an i.i.d.\ copy of a random variable $Y$, whose generating function is given by:
\begin{equation}
    P_Y(s) = \sum_{n=0}^\infty {\mathrm{Pr}}(Y = n)s^n = \frac{1}{\beta + \gamma}\left(\beta s^2 + \gamma\right),
\end{equation}
so that at each infection event an infectious case produces an identical copy of itself as well as an expected number of secondary infections given by
$$
\mathbb{E}[Y] -1 = 
\left.\frac{{\mathrm{d}}P}{{\mathrm{d}}s}\right|_{s=1} - 1 = \frac{\beta - \gamma}{\beta+\gamma}.
$$
This is equivalent to a birth-death chain model in which the expected number of secondary cases due to a single infected individual over the course of their infection is given by $R_0 = \frac{\beta}{\gamma}$ \cite{athreya2004branching, dorman2004garden}. 

The number of infectious cases at time $t$, $Z(t)$ then has the generating function $Q(t, s)$ given by:
\begin{equation}
    Q(t, s) = \mathbb{E}\left[s^{Z(t)} \right] = \sum_{n=0}^\infty \mathrm{Pr}(Z(t) = n)s^n . \label{Qdef}
\end{equation}
We obtain the probability of extinction for an outbreak starting with a single infectious case, which we denote $q(t)$, by setting $s=0$ in (\ref{Qdef}). It was shown by Harris \cite{harris1963theory} that $Q(t, s)$ satisfies the Chapman-Kolmogorov backward equations:
\begin{align} \nonumber
    \frac{\partial Q}{\partial t} &= -\rho \left[ Q(t, s) - P_Y(Q(t, s)) \right]
 = \beta Q^2 - \rho Q + \gamma , \\
     \quad Q(0, s) & = s,\label{Qeq}
\end{align}
where $\rho = \beta + \gamma$. 

 Solving (\ref{Qeq}) for $Q(t, s)$ and setting $s=0$ with initial condition $q(0) = 0$ gives an explicit expression for the extinction probability $q(t)$, given by:
\begin{equation}
    q(t) := Q(t, 0) = 1 - \left(\int_0^t \beta {\mathrm{e}}^{(\beta - \gamma)u} {\mathrm{d}}u +  {\mathrm{e}}^{(\beta - \gamma)t} \right)^{-1}, \label{Qsol}
\end{equation}
with details given in Appendix \ref{q_solve}. It is also straightforward to obtain the first and second moments, $m_1(t)$ and $m_2(t)$ of $Z(t)$ by differentiating (\ref{Qeq}) with respect to $s$ and substituting $s=1$ \cite{athreya2004branching}, so that:
\begin{align}
    \frac{{\mathrm{d}}m_1}{{\mathrm{d}}t} &= r m_1 & & \Rightarrow \quad m_1(t) = {\mathrm{e}}^{rt} \label{BPmean} \\
    \frac{{\mathrm{d}}m_2}{{\mathrm{d}}t} &= 2\beta (m_1)^2 + r m_2 & & \Rightarrow \quad m_2(t) = \frac{\beta}{r}\left({\mathrm{e}}^{2rt} - {\mathrm{e}}^{rt}\right) , \label{BPsecondmoment}
\end{align}
where $r = \beta - \gamma$ is the growth rate for the number of cases. The variance (i.e.\ second \emph{central} moment) at time $t$, $\sigma^2(t)$, is given by $\sigma^2(t) = m_2(t) - (m_1(t))^2$. 

\subsection{Time to Establishment of an Outbreak} \label{Tstardef}

We define the time at which an outbreak that begins with an initial case is fully established in the resident population, provided that it has not gone extinct, which we denote by the random variable $T$. Once this time has been reached, we conclude that the subsequent disease dynamics are well approximated by a deterministic model. In order for this to be the case, we require two conditions for $t>T$, namely that the local epidemic is growing approximately according to the mean growth curve and that the probability of having no cases is approximately constant. 

We choose an appropriate threshold $T^*$ based on these two conditions in order to find a distribution of the random variable $T$ that is centred on the threshold $T^*$, i.e. $T^* = {\mathbb{E}}[T]$. We formalise our criteria for choosing $T^*$ as follows:
    \begin{enumerate}
        \item $c(t) := \sigma(t)/m_1(t) = l +  \varepsilon_1$ for $\varepsilon_1 > 0$ and $t>T_1$ \label{cond1}
        \item $q(t)$ := $P(Z(t) = 0) =  q - \varepsilon_2$ for $0 < \varepsilon_2 < q$ and $t>T_2$, \label{cond2}
    \end{enumerate}  
where $c(t)$ is the coefficient of variation, $q = \lim_{t \to \infty} q(t)$ and $l$ is constant. We then choose $T^* = T^*(\varepsilon_1, \varepsilon_2)$ such that both of these conditions are satisfied, i.e. $T^* = {\mathrm{max}}\{T_1, T_2\}$. Note that in a supercritical process, corresponding to $R>1$, both of these conditions are guaranteed to be satisfied as $t \to \infty$ (see, for example, \cite[Ch.\ 7]{athreya2004branching} for details).

We interpret $T^*$ as the central estimate for $T$, the time at which the pathogen is established in the resident population and investigate the uncertainty around this estimate by obtaining an approximation to the first-passage time distribution for the branching process to the level $Z^* := m_1(T^*) = \mathbb{E}[Z(T^*)]$. 

\subsection{Diffusion Model Approximation} \label{FPT_Feller}
In order to make progress on a first-passage time distribution for our branching process, we first rely on a diffusion approximation studied by Kurtz \cite{kurtz1970solutions, kurtz1971limit} and Jagers \cite{jagers1971diffusion}. This diffusion approximation was first introduced by Feller \cite{feller1951genetics, feller1951}, and proved using the Fokker-Planck equations by Ji\v{r}ina \cite{jivrina1969}. This diffusion is a special case of the Kramers-Moyal expansion of the Kolmogorov equation for the branching process, which takes the first two terms of the Taylor expansion of the probability density function for the transition rates of the process \cite{van1992stochastic}. The Feller diffusion is also known as the squared Bessel process \cite{peskir2022sticky, peskirsticky} or, in finance, as the Cox-Ingersoll-Ross diffusion \cite{cox2005theory} and has been used extensively to study changes in interest rates. 

We begin with the classic Feller branching diffusion approximation to the single-type branching process. If $Z_t$ is the number of infectious cases at time $t$, then as $t$ grows large $Z_t$ obeys the following stochastic differential equation:
\begin{equation} 
    {\mathrm{d}}Z_t = r Z_t {\mathrm{dt}}+ \sqrt{\rho Z_t} {\mathrm{d}}W_t ,\label{FellerSDE}
\end{equation}
where $W_t$ denotes a standard Wiener process. As Feller noted in his original paper, the above process has an absorbing boundary at $Z_t = 0$, so there is a non-zero probability that the process becomes extinct. We can represent this SDE equivalently as a partial differential equation via the Fokker-Planck equation, so that:
\begin{equation}
    \frac{\partial f(t, x)}{\partial t} = \frac{\rho}{2} \frac{\partial^2}{\partial x^2} \left[x f(t, x) \right] - r \frac{\partial}{\partial x} \left[x f(t, x) \right] , \label{FellerPDE}
\end{equation}
where $f(t, x)$ is the probability density function of the number of cases at time $t$. Feller demonstrated that the solution is the PDF of a non-central $\chi^2$ distribution with zero degrees of freedom \cite{feller1951, siegel1979noncentral}, which has the explicit form:
\begin{equation}
    f(t, x) = \frac{2 r {\mathrm{e}}^{rt}}{\rho({\mathrm{e}}^{rt} - 1)} \sqrt{\frac{{\mathrm{e}}^{rt}}{x}} I_1 \left( \frac{4r\sqrt{x{\mathrm{e}}^{rt}}}{\rho({\mathrm{e}}^{rt} - 1)} \right) {\mathrm{exp}}\left(- \frac{r ({\mathrm{e}}^{rt}+x)}{\rho({\mathrm{e}}^{rt} - 1)}\right) , \label{chisq_pdf}
\end{equation}
where $I_1(\cdot)$ is the modified Bessel function of the first kind. It is possible to derive from first principles the PDF in \eqref{chisq_pdf} from \eqref{FellerPDE} using the method of characteristics (see Appendix \ref{Feller_solve}).

Note that \eqref{chisq_pdf} does not represent a true density, since $\int_0^\infty f(t, x) \, {\mathrm{d}}x = 1 - {\mathrm{exp}}\{-\lambda/2\} < 1$, where $\lambda = 4{\mathrm{e}}^{rt} / (\rho ({\mathrm{e}}^{rt} - 1)) $. This is because the non-central $\chi^2$ distribution has a mass at zero equal to ${\mathrm{exp}}(-\lambda/2)$, which represents the extinction probability at time $t$ for the Feller process. In order to obtain a true density for the number of cases at time $t$, we condition our process on never reaching zero cases to obtain a density $\hat{f}(t, x)$ defined by:
\begin{equation}
    \hat{f}(t, x) = \frac{f(t, x)}{(1-{\mathrm{e}}^{-\lambda/2})} .
\end{equation}
From the non-central $\chi^2$ distribution we obtain the First Passage Time (FPT) distribution for the Feller process, which approximates the FPT for the branching process. At each time $t$, the integrated density $\hat{F}(t, x) = \int_0^\infty \hat{f}(t, x) \, {\mathrm{d}}x = {\mathrm{Pr}}(Z_t \leq x)$ gives the probability that the number of cases has not yet reached the level $x$, given that the process has not yet become extinct. We then obtain the cumulative density function, $U_x(t) = {\mathrm{Pr}}(T_x \leq t)$ for the random variable $T_x = {\mathrm{inf}}\{t : Z_t \geq x\}$ directly from the CDF $\hat{F}(t, x)$ \cite{ditlevsen2008}. Choosing the level $x = Z^*$ with $Z^* = \mathbb{E}[Z(T^*)]$ defined in \S{}\ref{Tstardef} above for the branching process, and henceforth letting $T := T_{Z^*}$ and $U(t) := U_{Z^*}(t)$ for notational compactness, we have that:
\begin{equation}
    U(t) = 1 - \hat{F}(t, Z^*) = {\mathrm{Pr}}(T^* \leq t) .
\end{equation}

\subsection{Gaussian Process Approximation to the Feller Diffusion} \label{FPT_GP}

As well as using the non-central $\chi^2$ solution of \eqref{FellerPDE}, we can also approximate the FPT distribution of the Feller diffusion using a Gaussian Process. Using a Gaussian Process approximation instead of the exact non-central $\chi^2$ distribution for the Feller Diffusion has the advantage that such approximations are likely to be available for a more general set of diffusion problems \cite{kurtz1971limit}. In particular, Gaussian approximations to solutions of stochastic differential equations can be applied to diffusion problems with more than one type, as well as to time-inhomogeneous diffusion problems \cite{buckingham2018gaussian, archambeau2007gaussian}.

Archambeau et al. \cite{archambeau2007gaussian} note that a stochastic differential equation of the form:
\begin{equation}
    {\mathrm{d}}X_t = (A(t)X_t + b(t)){\mathrm{d}}t + \sqrt{V(t)} {\mathrm{d}}W_t \label{GP_SDE}
\end{equation}
has a Gaussian Process solution GP($m(t), \Sigma(t)$), whose mean and variance satisfy the following ordinary differential equations:
\begin{align}
    \frac{{\mathrm{d}}m}{{\mathrm{d}}t} &= A(t) m(t) + b(t) , \label{GPmean}\\
    \frac{{\mathrm{d}}\Sigma}{{\mathrm{d}}t} &= 2A(t)\Sigma(t) + V(t) . \label{GPvar}
\end{align}
This Gaussian Process solution does not require the process to be homogeneous in time and also has an analogous formulation if $X_t, b(t)$ and ${\mathrm{d}}W_t  \in {\mathbb{R}}^n$ are vectors and $A(t), V(t) \in {\mathbb{R}}^{n\times n}$ are matrices. The solutions of \eqref{GPmean} and \eqref{GPvar} are chosen such that the Kullback-Leibler divergence between the distribution of $X_t$ and the Gaussian Process GP($m(t), \Sigma(t)$) is minimised (see \cite{archambeau2007gaussian}).

In order to make a Gaussian Process approximation to the Feller diffusion, we first need to transform the Feller stochastic differential equation \eqref{FellerSDE} using It\^{o}'s Lemma \cite{ito1951stochastic} so that it has the form \eqref{GP_SDE}. We start by making the transformation $X_t = h(t, Z) = \sqrt{Z_t} $, so that \eqref{FellerSDE} becomes:
\begin{align}
    {\mathrm{d}}X_t &= \left(\frac{r}{2}Z_t \frac{\partial h}{\partial Z} + \frac{\rho Z_t}{2} \frac{\partial^2 h}{\partial Z^2}\right){\mathrm{d}}t + \sqrt{\rho Z_t} \frac{\partial h}{\partial Z} {\mathrm{dW}}_t \nonumber \\
    &= \left(rX_t - \frac{1}{8X_t}\right){\mathrm{d}}t + \frac{\sqrt{\rho}}{2} {\mathrm{dW}}_t . \label{sqrt_full}
\end{align}
From the above equation, we make the approximation that $\mathcal{O}(X_t^{-1})$ terms are small, and can therefore be ignored, so that our equation finally becomes:
\begin{equation}
    {\mathrm{d}}X_t \approx rX_t {\mathrm{d}}t + \frac{\sqrt{\rho}}{2} {\mathrm{dW}}_t, \label{Feller_sqrt}
\end{equation}
which is in the form of \eqref{GP_SDE}. This approximation does not hold when $X_t \ll 1$, since for small $X_t$ the original process $Z_t$ is close to extinction. We should also note that \eqref{Feller_sqrt} no longer has $X_t=0$ as an absorbing state. We must therefore impose the additional boundary restriction $X_t \geq 0$ and ${\mathrm{d}}X_t = 0$ for $X_t = 0$ in order for the process \eqref{Feller_sqrt} to share the same properties as \eqref{FellerSDE}. Now that we have our SDE in the required form, we can write a Gaussian Process solution for our approximation GP($m(t), \Sigma(t)$), for which we solve the ODEs:
\begin{align}
    \frac{{\mathrm{d}}m}{{\mathrm{d}}t} &= rm(t) \label{GPmean_Feller} ,\\
    \frac{{\mathrm{d}}\Sigma}{{\mathrm{d}}t} &= 2r \Sigma(t) + \frac{\rho}{4} . \label{GPvar_Feller}
\end{align}
As with the Feller diffusion, we can now obtain the first passage time distribution for the Gaussian Process hitting the level $\sqrt{Z^*}$ directly from the CDF, $\Phi(\cdot)$, of the Gaussian distribution at each time $t$, this time conditioning on the process being greater than 0 (since the Gaussian Process at time $t$ may also take negative values, unlike the non-central $\chi^2$ distribution). If $U_G(t)$ is the PDF for the first passage time distribution of the Gaussian Process to the level $\sqrt{Z^*}$, conditional on the process being greater than zero, then we have the explicit expression:
\begin{equation}
    U_G(t) = \frac{1 - \Phi(\sqrt{Z^*} \, ; \, m(t), \Sigma(t))}{1 - \Phi(0\, ; \, m(t), \Sigma(t))} . \label{FPT_GP_dist}
\end{equation}

\subsection{Linear Noise Approximation}
The Linear Noise Approximation (LNA) is a standard method used to approximate solutions of stochastic differential equations introduced by van Kampen \cite{kampen1961power}. We consider the LNA to Equation \eqref{FellerSDE} as an additional comparison to the approximations that we obtain in the previous sections. The LNA is based on rewriting the stochastic process $Z_t$ as the sum of a deterministic part, $\varphi(t)$ and a stochastic noise term $\xi_t$. Choosing $\varphi(t) = {\mathrm{e}}^{rt}$ to be the deterministic part, we make the substitution $Z_t = {\mathrm{e}}^{rt} + \xi_t$, so that:
\begin{align}
    {\mathrm{d}}Z_t &= r{\mathrm{e}}^{rt} {\mathrm{d}}t + {\mathrm{d}}\xi_t \nonumber \\
     &=  r Z_t {\mathrm{dt}}+ \sqrt{\rho Z_t} {\mathrm{dW}}_t  \nonumber \\ 
     \Rightarrow  {\mathrm{d}}\xi_t &= r \xi_t {\mathrm{dt}} + \sqrt{\rho {\mathrm{e}}^{rt} (1 + {\mathrm{e}}^{-rt} \xi_t)} {\mathrm{dW}}_t .
\end{align}
Based on the final expression for ${\mathrm{d}}\xi_t $ and, assuming ${\mathrm{e}}^{-rt} \xi_t$ to be small, we may use a power series expansion for $\sqrt{1 + {\mathrm{e}}^{-rt} \xi_t}$ in order to obtain the first order Linear Noise Approximation
\begin{equation}
    {\mathrm{d}}\xi_t \approx r \xi_t {\mathrm{dt}} + \sqrt{\rho} {\mathrm{e}}^{rt/2} {\mathrm{dW}}_t \label{LNA1_eq}\; ,  
\end{equation}
where in the power series we have ignored terms that are $\mathcal{O}(\xi_t{\mathrm{e}}^{-rt/2})$. We note that this is not a well-controlled
expansion since this quantity will not typically be small compared to $1$, however, it is
included since such expansions are an extremely popular approach and may be attempted without guarantees of convergence.

As with the square root of the Feller process, equation \eqref{LNA1_eq} can be solved using equations \eqref{GPmean} and \eqref{GPvar} in order to give a solution that is normally distributed. Solving in this way, we find that the first-order Linear Noise Approximation to $Z_t$ is given by a Gaussian distribution that takes the form:
\begin{equation}
    Z_t \approx \mathcal{N}\left({\mathrm{e}}^{rt}, \frac{\rho}{r}({\mathrm{e}}^{2rt} - {\mathrm{e}}^{rt})\right). \label{LNA_Gaussian}
\end{equation}

\subsection{Peak Time Distribution for a Deterministic Model} \label{Peak}

The distributions of first passage times obtained in the previous section induce a distribution on the time taken for the resulting outbreak to hit its peak. Once the threshold $Z^*$ is reached, we model the subsequent epidemic using the standard deterministic SIR model of Kermack and McKendrick \cite{kermack1927contribution} assuming constant infectiousness of exponential duration. We consider a closed population of size $N$, with an initial number of infectious cases $Z^*$ that starts at time $t = T^*$, giving the ordinary differential equation system: 
\begin{align}
    \frac{{\mathrm{d}}S}{{\mathrm{d}}t} & = - \frac{\beta S I}{N} , \nonumber \\
    \frac{{\mathrm{d}}I}{{\mathrm{d}}t} & = \frac{\beta S I}{N} - \gamma I , \nonumber \\
    S(T^*) & = N - Z^* , \nonumber \\  
    I(T^*) & = Z^* ,\nonumber \\
    R(t) & = N - S(t) - I(t) , \quad \forall t \in [T^*,\infty) ,
\label{SIR_eqs}
\end{align}
which we can solve using standard numerical integration routines as an initial value problem for $t \in [T^*,\infty)$.
We have implicitly assumed that $R(T^*) \approx 0$ since, for the branching process, we assume that the number of susceptible individuals in the population is not significantly depleted so that $S(T^*) \approx N$. We note that this assumption is also required up to time $t=T^*$ in order for the linear branching process to be a valid approximation of the epidemic dynamics. However, this observation suggests a trade-off in the choice of $\epsilon$ and, hence, of $Z^*$. The choice of threshold $\epsilon$ should not be so small as to make $Z^*$ large enough that the assumption of negligible depletion of the susceptible population is no longer valid. One could improve upon this assumption by considering the total progeny of the branching process in order to keep track of individuals that have been infected but have since recovered, but we find that making this assumption does not have a large impact on our results. A comparison of the true peak timing for the stochastic SIR epidemic compared with the estimated peak timing using a hybrid branching process and deterministic model is given in Appendix \ref{truepeak_timing}. 

Solving \eqref{SIR_eqs}, we can obtain the time at which the epidemic reaches its peak, $t_{\text{peak}}$. We can then simply take the distribution of the hitting time for the peak of the epidemic to be the same as the FPT distribution centred on $T^*$ obtained in the previous sections, translated forwards by the difference $t_{\text{peak}} - T^*$. This is equivalent to the simulated distribution that we would expect to obtain if we ran multiple epidemics by solving \eqref{SIR_eqs} and drawing random initial times $t_0$ from the FPT distribution $T$.

\subsection{Extension to a model with Heterogeneous Susceptibility}

So far, we have considered only single-type, Markovian epidemics. However, many extensions are possible and have been discussed extensively by other authors (see, for example, \cite{novozhilov2008spread} and \cite{diekmann2013mathematical}). 
One extension that could be incorporated straightforwardly within the formalism introduced in the previous sections is the model where individuals have heterogeneous susceptibility, but are otherwise identical in terms of their infectivity, and are assumed to mix homogeneously. 
Models where only the susceptibility is assumed to be heterogeneous have been applied to COVID-19, for example in the work by Gomes et al. \cite{gomes2022individual} (although they assumed an SEIR model structure). 
Whether the heterogeneity in susceptibility is modelled by subdividing the population in a discrete number of classes or drawing each individual's susceptibility from a known distribution, the branching process formalism developed here is sufficient because the dynamics of $I$ can be represented by a system identical to that of the single-type Markovian model, for a suitably altered version of the transmission rate $\beta$. 
To see this, consider Equation (2) in \cite{novozhilov2008spread}:
\begin{equation}
    \frac{\mathrm{d}}{\mathrm{d}t}I(t) = \Bar{\beta}(t)S(t)I(t) - \gamma I(t),
\label{Nov_dyn}
\end{equation}
where $\Bar{\beta}(t) = \int_\Omega \beta(\omega) p_s(t,\omega)\mathrm{d}\omega$, with $\beta(\omega)$ the transmission rate towards an individual with susceptibility indicated by $\omega$ and $p_s(\omega)$ the probability density function for the susceptibility $\omega$ of an individual chosen uniformly at random from the population, i.e.\ $p_s(t, \omega) = s(t,\omega)/S(t)$ with $s(t,\omega)$ being the density of individuals with susceptibility $\omega$ at time $t$. 

In the linearised system describing the early epidemic dynamics, $S(t)\equiv 1$, so that $S, s$ and $p_s$, and hence $\Bar{\beta}$ are independent of time, leading to
\begin{equation}
    \frac{\mathrm{d}I}{\mathrm{d}t} = \Bar{\beta} I - \gamma I,
\end{equation}
the deterministic limit of a simple birth-death process with infection rate $\Bar{\beta}=\int_\Omega \beta p_s(\omega) \mathrm{d}\omega$ rather than simply $\beta$.




Beyond the early branching process approximation, Equation \eqref{Nov_dyn} can be used as the deterministic limit to calculate the actual time to the peak (note that, when the full non-linear dynamics are considered, this model differs from the single-type Markovian SIR model, because more susceptible individuals get ``burnt out'' more quickly). An even simpler form for the equation could be derived in special cases. For example, Novozhilov \cite{novozhilov2008spread} showed how, if the susceptibility, $\omega$, of individual $j$ was drawn (independently from each other individual) from a Gamma distribution, i.e.\ $\omega \sim \Gamma(r, k)$, with rate parameter $r$ and shape parameter $k$, then the stochastic model
converges to a deterministic approximation given by the following system of ODEs:
\begin{equation}
     \frac{\mathrm{d}S}{\mathrm{d}t} = -\beta \left(\frac{S}{N}\right)^{\theta + 1} I ; \qquad
     \frac{\mathrm{d}I}{\mathrm{d}t} = \beta \left(\frac{S}{N}\right)^{\theta + 1} I  - \gamma I ,
\end{equation}
where $\theta=k^{-1}$. Similar closed-form expressions can be derived for other distributions of susceptibility, such as a 
Wald (Inverse Gaussian) or Weibull distribution -- 
see \cite{novozhilov2008spread} for more examples and special cases.

Further extensions are not captured directly with the formalism presented in the previous sections, but are still possible with a suitable adaptation of the methodology presented here, which is a matter of ongoing work. We conjecture that it should be possible to provide suitable approximations for the First Passage Time distribution in much more general cases, for example for a model where each susceptible $j$ in a population $\mathcal{N}$ experiences a \emph{force of infection}  
given by:
\begin{equation}
\Lambda_j = \sum_{i \in \mathcal{N}} \lambda(t - T_i, A_i, A_j) ,
\end{equation}
where $\lambda$ is the force of infection that individual
$i$ exerts on individual $j$, which could depend also on the time elapsed since $T_i$, the time of infection of $i$, and $A_i$ and $A_j$ are the types of individuals $i$ and $j$, respectively, which belong to a suitable set of types $\mathcal{A}$. 
Our conjecture relies on the fact that this sort of general model has, under certain technical conditions, a known deterministic limit \cite{barbour2013approximating} and branching process approximation that could be formulated using Bellman-Harris or Crump-Mode-Jagers processes, to allow non-exponential recovery rates \cite{barbour2013approximating, haccou2005branching}. 
However, to our knowledge, diffusion limits are known only in certain specific cases, and a completely general diffusion limit for non-Markovian stochastic models is still lacking, so each extension of our model would require additional work and a setting-specific approach. Additionally, one could discretise any continuous heterogeneity with a discrete population risk or contact structure, which one could analyse using a multi-type branching process model \cite{dorman2004garden}. We also do not consider changes in the population size or immigration of infectious cases, though these have been considered elsewhere \cite{ball2017epidemic}.

\section{Results}

In order to compare the FPT distributions obtained in \S{}\ref{FPT_Feller} and \S{}\ref{FPT_GP} above we model an outbreak of an infectious disease using our SIR branching process model defined in \eqref{BP_outbreak}. As a baseline, we take $R_0 = 2$ with an infectious period of 7 days so that $\beta = 2/7$ and $\gamma = 1/7$. These values correspond to a doubling time of $t_D = 4.85$ days, which is quantitatively not dissimilar from some early estimates of the doubling time of COVID-19 in China \cite{pellis2021challenges}. However, we have also considered other values of $\beta$ and $\gamma$ to check the sensitivity of our results to different parameter choices. We also consider a closed population of size $N = 10^7$, so as to model an outbreak in a population the size of a large city similar, for example, to London, UK. 

\subsection{Branching Process SIR model}
\begin{figure}
    \centering
    \includegraphics[scale = 0.425]{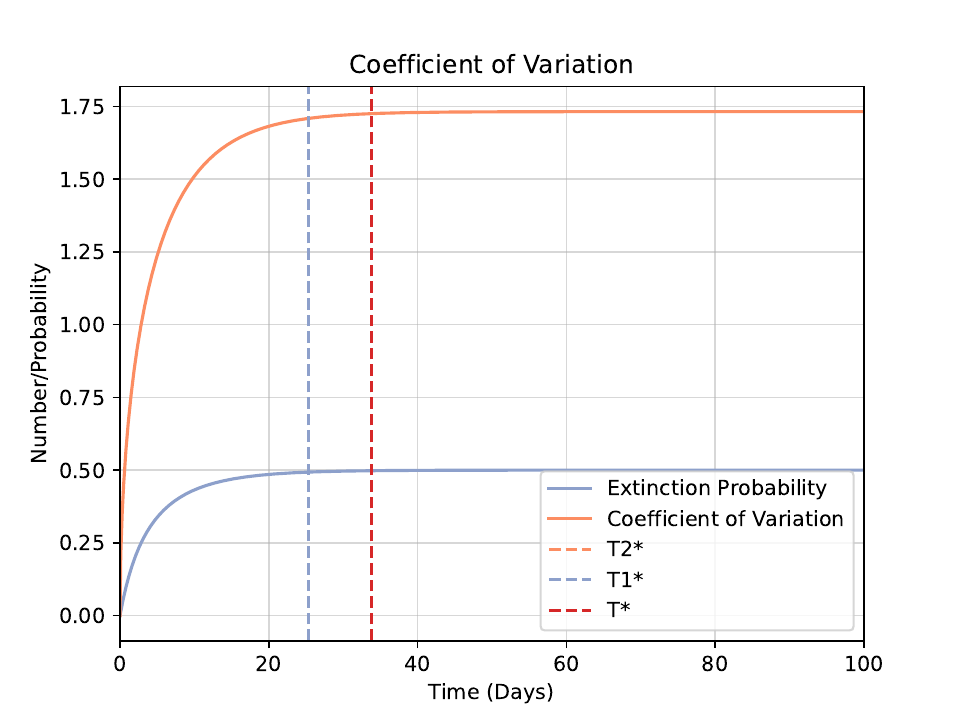}
    \includegraphics[scale = 0.5]{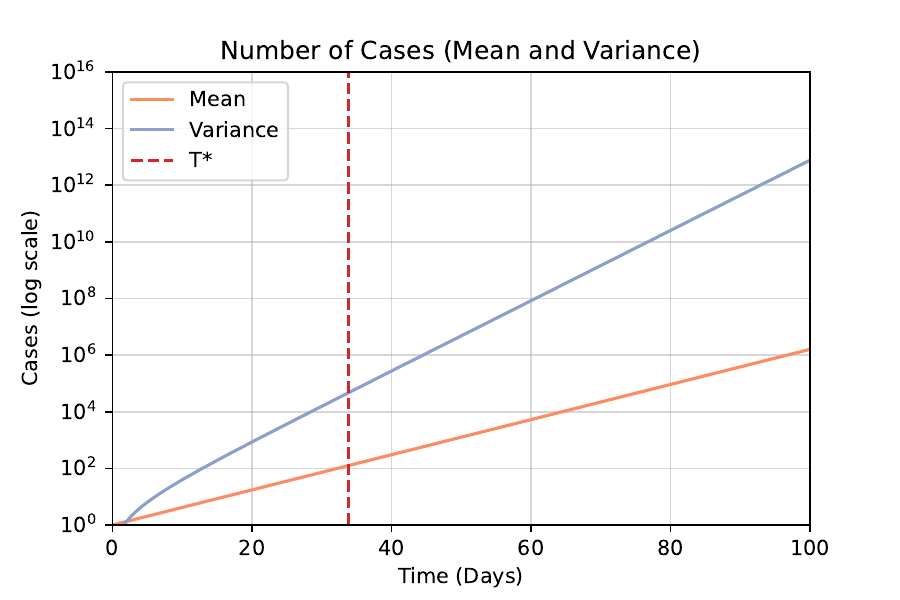}
    \caption{Extinction probability, $q(t)$, and coefficient of variation, $c(t) = \sigma(t)/m_1(t)$, over time, for $R = 2$. From these two curves, we choose the time $T^*$ at which both are within $\epsilon$ of their asymptotic limits. For $\epsilon = 10^{-3}$, we have $T^* = 34$ days. The mean number of cases for the branching process at time $t = T^*$ is given by $\mathbb{E}[Z(T^*)] = 125$ cases.}
    \label{fig:Tstar}
\end{figure}

In order to choose the time at which we are able to switch from a stochastic model to a deterministic one, we calculate \eqref{Qsol} for our branching process model together with \eqref{BPmean} and \eqref{BPsecondmoment} in order to obtain $q(t)$ and $c(t)$ for our choices of parameters. These results are plotted in Figure \ref{fig:Tstar}. We choose a common threshold $\epsilon$, defined in Section \ref{Tstardef}, for both curves in order to calculate the times $T_1^*$ and $T_2^*$  after which $q(t)$ and $c(t)$ are approximately constant, respectively. Figure \ref{fig:Tstar} shows the resulting choice for $T^*$ obtained by taking the maximum of these two times. For our baseline model, with $R=2$, we choose $\epsilon = 10^{-3}$ and obtain $T^* = 34$ days, with a mean of $Z^* = \mathbb{E}[Z(T^*)] = 125$ cases for the branching process. 

To investigate the true underlying FPT distribution for the branching process to the level $Z^* = 125$, we simulate sample trajectories of our branching process using the Gillespie algorithm \cite{gillespie1977exact}, stopping the algorithm at the time at which the number of cases reaches $Z^*$. We run $10^5$ simulations of the branching process, stopping each simulation once the number of cases reaches either zero or $Z^*$, and obtain an FPT distribution based on the stopping times for each simulation. We discard simulations for which the branching process goes extinct, so that the FPT distribution is conditioned on non-extinction. This ensures consistency with the approximations of the FPT made using the Feller Process and Gaussian Processes, which we also condition on the number of cases not reaching zero. We treat this sampled FPT distribution as a benchmark, to which we compare the distributions obtained via both the Feller and Gaussian Process approximations of the FPT distribution.

\subsection{Feller Diffusion and Gaussian Process Approximations}
\begin{figure}
    \centering
    \includegraphics[scale=1.0]{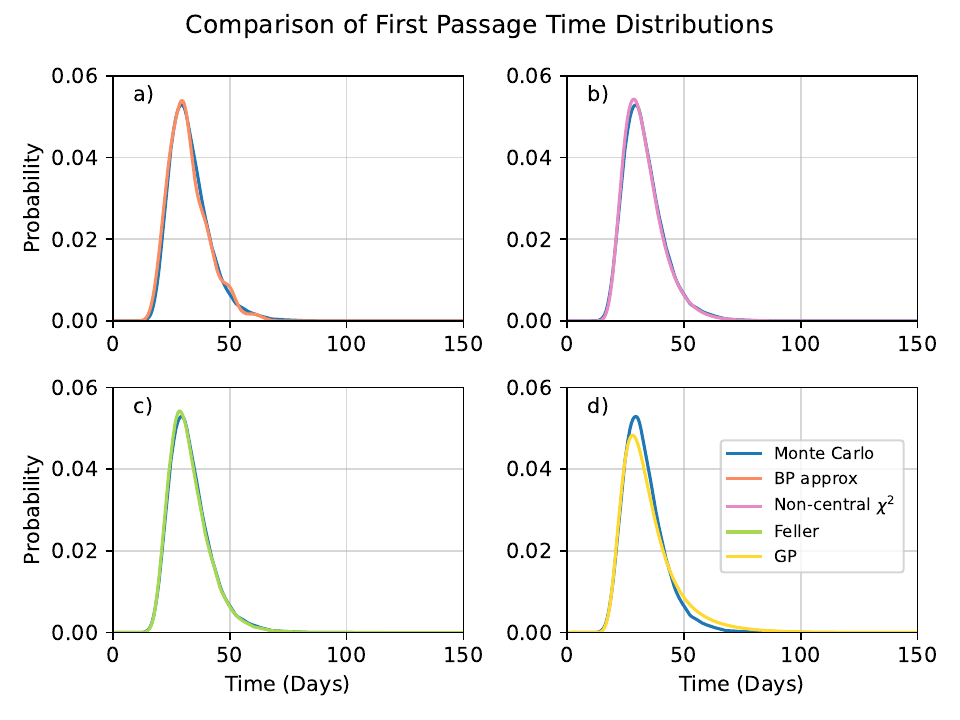}
    \caption{Comparison of estimated probability density functions for the First Passage Time distribution with for $R_0 = 2$ and $Z^* = 125$ cases, using the $10^5$ simulations of the branching process as a benchmark. We compare FPT distributions obtained via a) $10^3$ simulations of the branching process (labelled `BP approx.'), b) the exact non-central $\chi^2$ distribution for the Feller process, c) 100,000 simulations of the Feller process (labelled `Feller') and d) the exact distribution for our Gaussian Process approximation (labelled `GP').}
    \label{fig:FPT_comparison}
\end{figure}

To compare the FPT distribution for the Feller diffusion with that of the branching process, we make use of the analytic result \eqref{chisq_pdf} and compare this with simulations of the Feller process using the Euler-Maruyama method \cite{higham2001algorithmic}. Comparisons of both the simulated and analytic FPT distributions are shown in figure \ref{fig:FPT_comparison}. For simulating the Feller diffusion, we ran $10^5$ simulations and compared the resulting FPT distribution with the analytic distribution derived from the non-central $\chi^2$ distribution, and found that they are almost identical. We also note that, compared to simulating the branching process via the Gillespie $\tau$-leaping algorithm, running the Euler-Maruyama simulation required significantly lower total computation time, even for a ten-fold increase in the number of simulations (see Table \ref{tab:comp_time}). We also calculate the FPT distribution based on the Gaussian Process approximation described by Equation \eqref{FPT_GP_dist}.

In order to evaluate the performance of each of our approximations, we compare the cumulative density functions (CDFs) obtained by both the Feller and Gaussian Process approximations with the (empirical) CDF of our simulation output for the branching process. We measure the closeness of each distribution to the simulated ``true'' distribution using both the Kullback-Leibler divergence and the Kolmogorov-Smirnov distance. We also compare our results with the empirical CDF of the simulation output obtained with a lower number of simulations, in order to demonstrate the trade-off between accuracy and computational cost. A comparison of the PDF of the first passage time distribution, $T$, estimated using all of these methods is given in Figure \ref{fig:FPT_comparison}. A comparison of the required computation time for calculating the FPT distribution using each of the methods described in this paper is given in Table \ref{tab:comp_time}. 

\begin{figure}
     \centering
    \includegraphics[scale = 0.6]
    {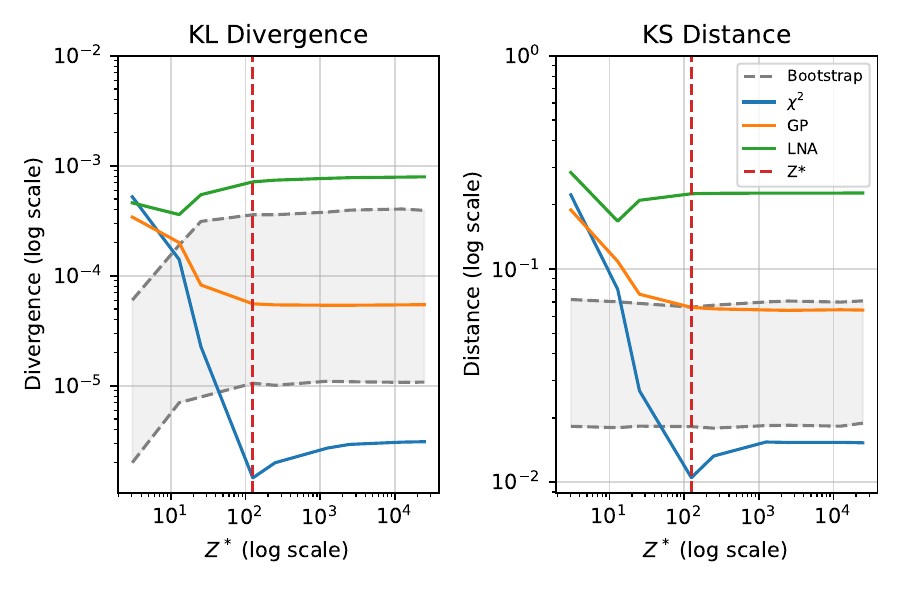}
    \includegraphics[scale=0.6]{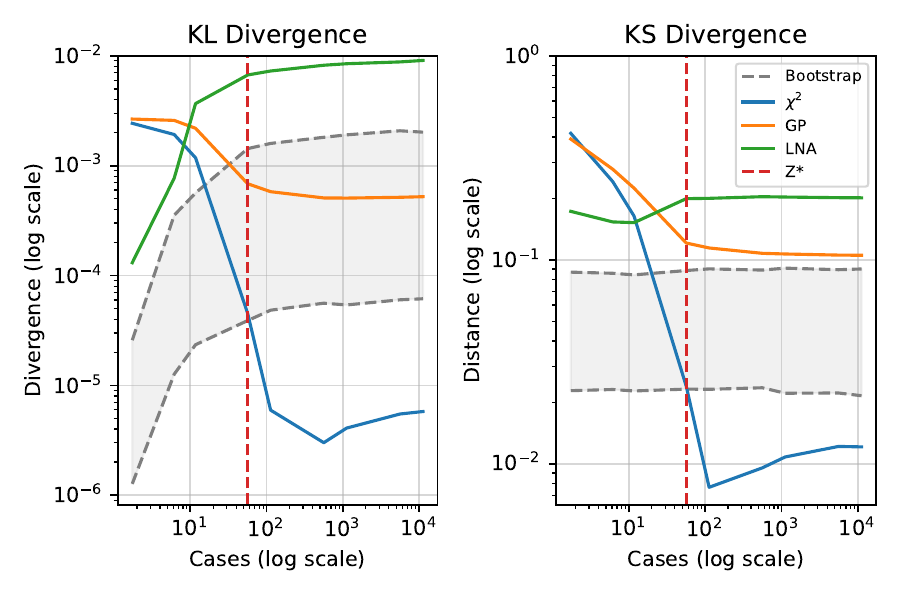}
    \includegraphics[scale=0.55]{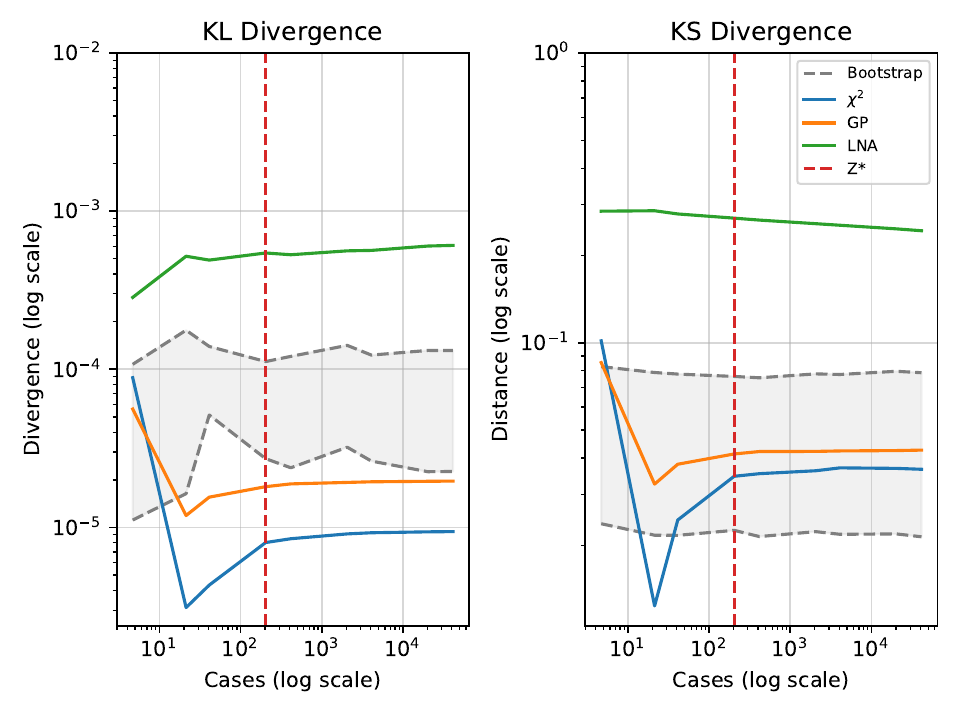}
     \caption{Convergence of approximated FPT distributions to the true distribution (estimated via $10^5$ simulations of the Gillespie algorithm) for an outbreak with $R_0 = 2$ (top), $R_0 = 1.5$ (middle) and $R_0 = 3$ (bottom), given different choices of $Z^*$. We compare approximations using the non-central $\chi^2$ distribution ($\chi^2$), the Gaussian Process approximation (GP) and the first-order Linear Noise Approximation (LNA) using the Kullback-Leibler divergence and the Kolmogorov-Smirnov distance between distributions. We also compare our approximations with the 95\ts{th} percentiles from $10^3$ bootstrapped samples of size $10^3$ of the branching process simulated using the Gillespie algorithm. The baseline $Z^*$, which corresponds to a choice of $\epsilon = 10^{-3}$ cases is plotted in red (dashed).}
     \label{fig:convergence}
 \end{figure}

In Figure \ref{fig:convergence}, we demonstrate the convergence of each approximated FPT distribution to the true distribution as the threshold $\epsilon$ is changed, corresponding to different choices for $T^*$ and for $Z^*$. We find that, of all of our methods, the non-central $\chi^2$ distribution arising from the Feller approximation to the branching process provides the closest approximation for almost all values of $\epsilon$ that we considered, both with respect to the Kullback-Leibler divergence and the Kolmogorov-Smirnov distance. The Guassian Process approximation of the FPT distribution also demonstrates good convergence with KL divergence below $10^{-4}$ for all choices of $Z^*$ above the baseline $Z^* = 125$. Both of our approximations perform significantly better than the first order Linear Noise Approximation with respect to both the KL divergence and the KS distance metric. 

\begin{table}
    \centering
    \begin{tabular}{|c|c|cc|cc|cc|}
        \hline & & \multicolumn{2}{|c|}{$\epsilon = 10^{-3}$} & \multicolumn{2}{|c|}{$\epsilon = 10^{-4}$} & \multicolumn{2}{|c|}{$\epsilon = 10^{-5}$} \\ 
         Method & Runs & Time & it/s & Time & it/s & Time & it/s  \\ \hline
         Gillespie & $10^5$ & 0:20:43 &  80.4 & 1:14:52 & 22.26 & 5:32:03 &  5.02 \\ 
         Euler-Maruyama & $10^5$  & 0:01:35 & 1045 & 0:02:13 & 746.79 & 0:04:29 & 370.79 \\ 
         Non-central $\chi^2$ & 1  & 0:00:25 & 0.04 & 0:00:25 & 0.04 & 0:00:25 & 0.04 \\
         Gaussian Process & 1  & 0:00:07 & 0.14 & 0:00:07 & 0.14 & 0:00:07 & 0.14 \\ \hline
    \end{tabular}
    \caption{Run times for each different method approximating the FPT distribution, for different choices of $\epsilon$. For large numbers of events, the Gillespie algorithm takes significantly longer to run than the other methods of estimating the FPT that rely on uniform time steps of size ${\mathrm{d}}t = 0.1$. Since the Non-central $\chi^2$ and Gaussian Process approximations are analytic, only one iteration is required.}
    \label{tab:comp_time}
\end{table}

\begin{figure}
    \begin{subfigure}[c]{0.3\textwidth}
        \centering
        \includegraphics[scale = 0.5]{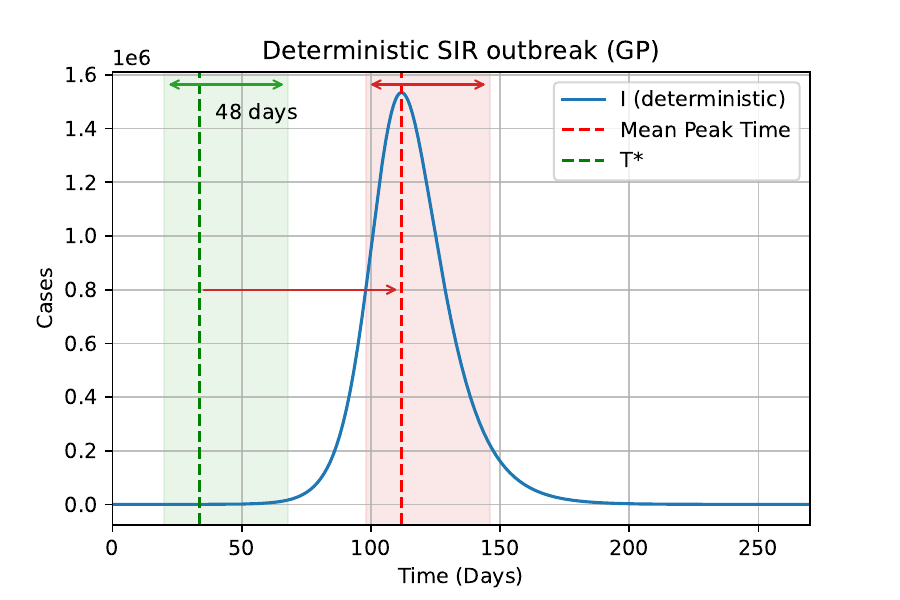}
    \end{subfigure}
    \hspace{2.5cm}
    \begin{subfigure}[c]{0.3\textwidth}
        \centering
        \includegraphics[scale = 0.5]{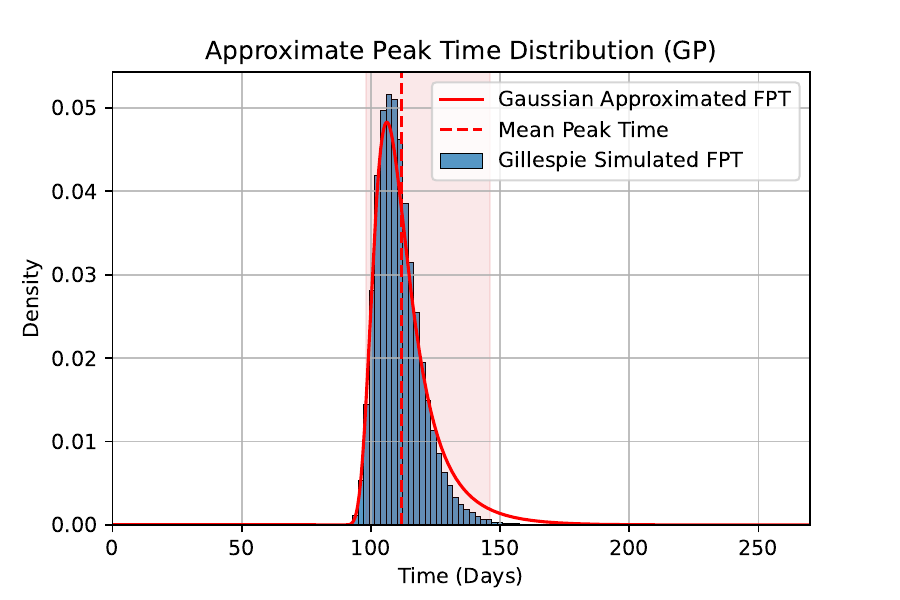}
    \end{subfigure}
    \vskip\baselineskip
    \begin{subfigure}[c]{0.3\textwidth}
        \centering
        \includegraphics[scale = 0.5]{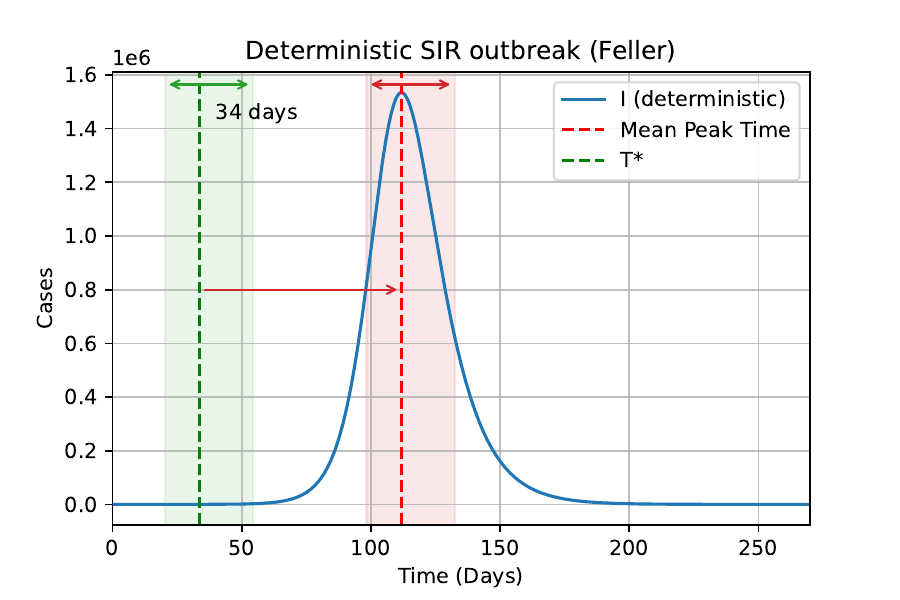}
    \end{subfigure}
    \hspace{2.25cm}
    \begin{subfigure}[c]{0.3\textwidth}
        \centering
        \includegraphics[scale = 0.5]{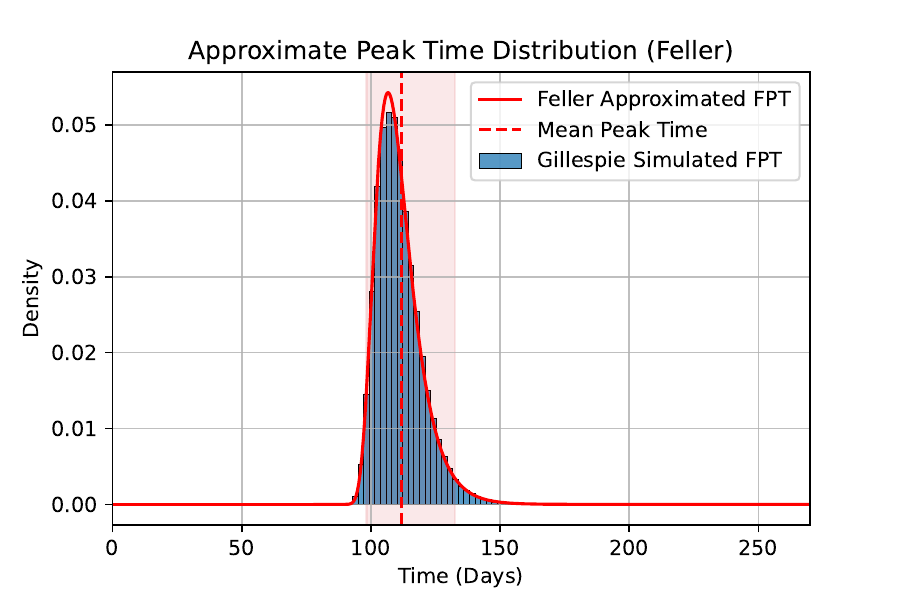}
    \end{subfigure}

    \caption{Deterministic outbreak for our baseline scenario with $R_0 = 2$ and $T^* = 34$ days. Starting from time $t_0 = T^*$, shown in green (dashed line), we solve the SIR equations for an outbreak with $I_0 = Z^*$ initial cases. We take the 5\ts{th} and 95\ts{th} percentile from the FPT distribution (green band) which we translate forward in time to obtain the uncertainty around the mean peak time (red band). Top left: Peak timing uncertainty of 48 days due to the Gaussian Process approximation. Top right: Peak time distribution for the Gaussian Process approximation, compared with the simulated distribution generated from the branching process. Bottom left: Peak timing uncertainty of 34 days due to the Feller Diffusion approximation.  Bottom right: Peak time distribution for the Feller diffusion approximation, compared with the simulated distribution generated from the branching process.}
    \label{fig:SIR_Delay}
\end{figure}
\subsection{Peak Timing Distribution} \label{peak_time}
Having obtained first passage time distributions for the number of cases to reach the level $Z^*$, we translate the distribution forward in time using the deterministic model described in Section \ref{Peak} so that the mean peak time coincides with the peak obtained from the deterministic SIR equations. The resulting peak time distributions using the Feller and Gaussian Process approximations are shown in Figure \ref{fig:SIR_Delay}. We also show the window of uncertainty around the peak in which 95\% of the distribution of peak times lie. This provides a useful tool for planning the allocation of resources and interventions required during the peak of an epidemic, including increasing hospital capacity \cite{vekaria2021}.

Relative to the FPT distribution obtained from the Feller diffusion approximation, we find that our Gaussian Process approximation has a longer-tailed FPT distribution. This accounts partly for the somewhat poorer convergence of the Gaussian Process approximation to the underlying distribution with respect to the Kolmogorov-Smirnov metric, and results in larger uncertainty in estimating the peak time. For $R_0=2$, we observe a 48-day window in which the peak is likely to fall using the Gaussian Process approximation, compared with a 34-day window for the equivalent Feller diffusion approximation. 

In order to provide sensitivity analysis for our results, we also demonstrate the convergence of the Feller and Gaussian Process approximations to the simulated Gillespie simulations of the FPT distribution for outbreaks with $R_0=1.5$ and with $R_0=3.$ For the outbreak with $R_0=1.5$, we also adjust the infectious period duration to 10 days, in order to show that our results can be obtained with different lengths of infectious period. These results are shown in Figure \ref{fig:convergence}.

We also obtain analogous figures for the peak timing distribution for different values of $R_0$ and with a different value of the recovery rate, $\gamma = 10^{-1}$. 
Comparing the results corresponding to different values of $R$ in Figure \ref{fig:convergence}, we see that our approximations achieve good convergence to the true underlying FPT distribution. The non-central $\chi^2$ distribution consistently outperforms the Gaussian Process approximation in terms of convergence, which reflects the fact that the Gaussian Process requires a further approximation of the square root of the Feller Process. We noted in Section \ref{peak_time} that the Gaussian Process approximation results in a fatter-tailed peak time distribution than for the true distribution based on the branching process approximation. This results in the Gaussian Process approximation performing worse with respect to the Kolmogorov-Smirnov metric than with respect to the Kullback-Leibler divergence. This also suggests that the Gaussian Process approximation captures the overall distribution reasonably well, but that it captures the shape of the tail less accurately than our other methods.

We also note that, whilst the non-central $\chi^2$ distribution provides a similar level of accuracy across all values of $R_0$ tested, the Gaussian Process approximation performs significantly better for higher values of $R_0$. With $R_0 = 1.5$, the lowest value of $R_0$ that we tested, the Kullback-Leibler divergence in the Gaussian Process is of order $10^{-3}$, which improves to an error of order $10^{-5}$ for $R_0 = 3$. This improvement in the KL divergence as $R_0$ increases is also reflected in the KS distance between the Gaussian Process and the true underlying distribution. 

\begin{figure}
    \centering
    \includegraphics[scale=0.4]{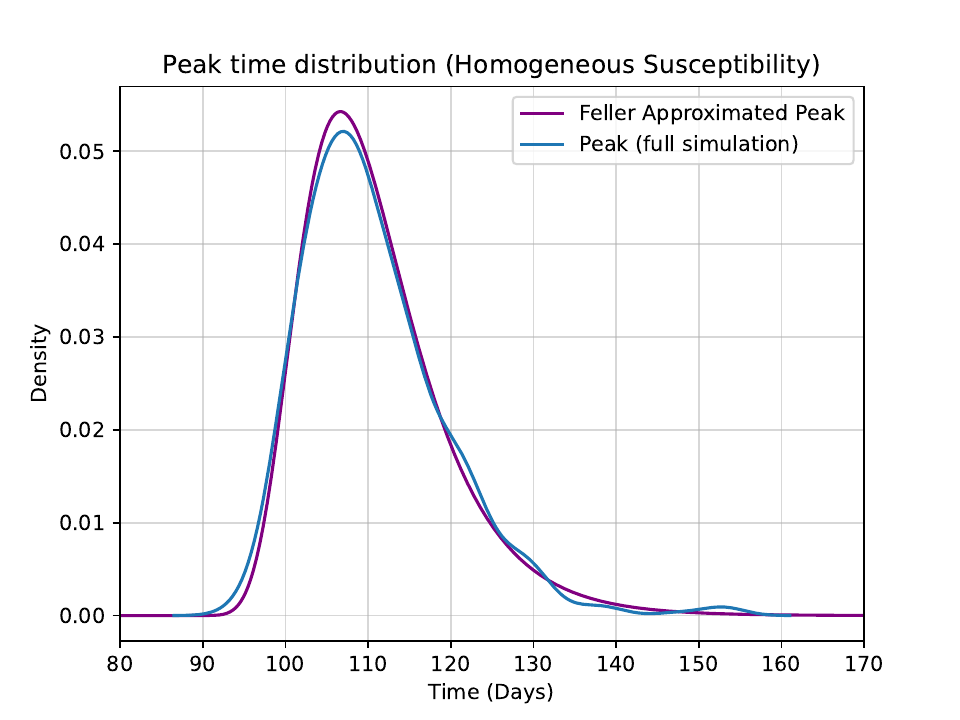}
    \includegraphics[scale = 0.4]{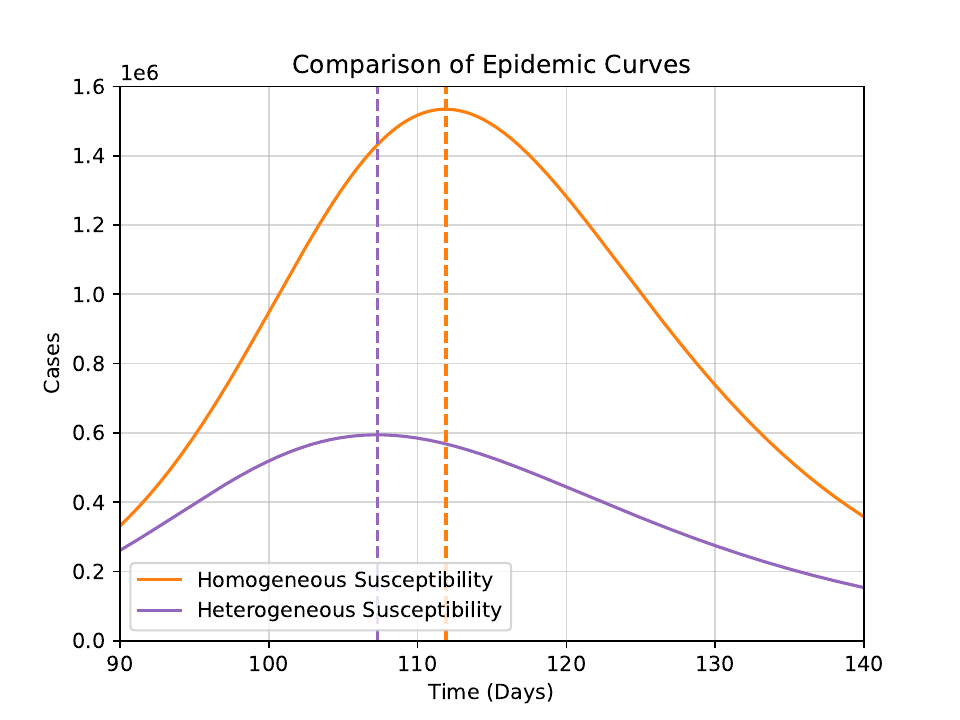}\\
    \includegraphics[scale = 0.4]{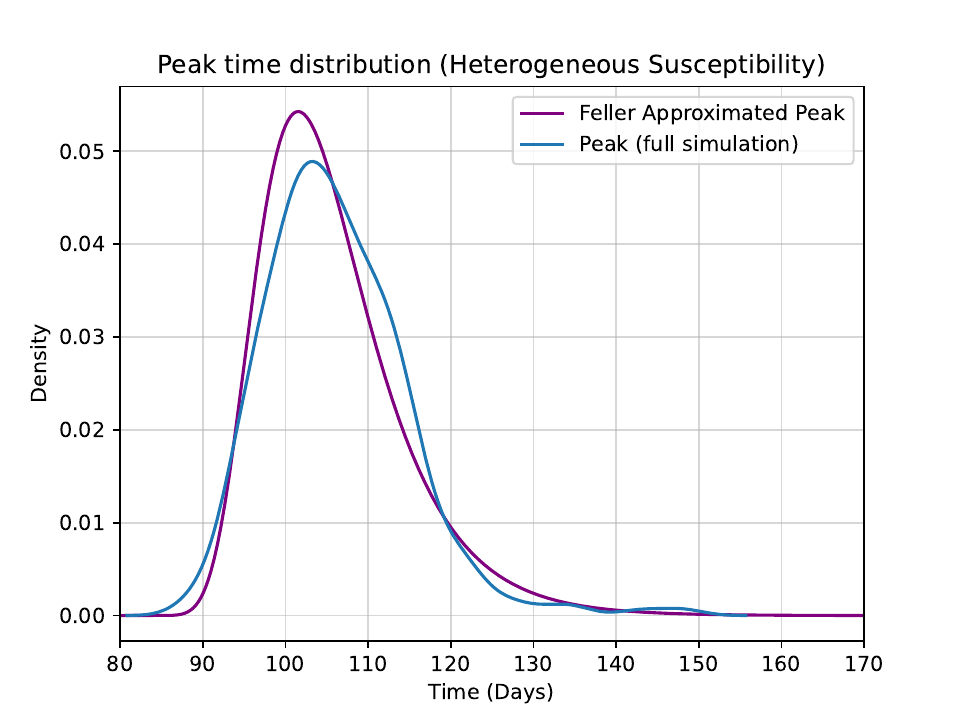}
    \includegraphics[scale = 0.4]{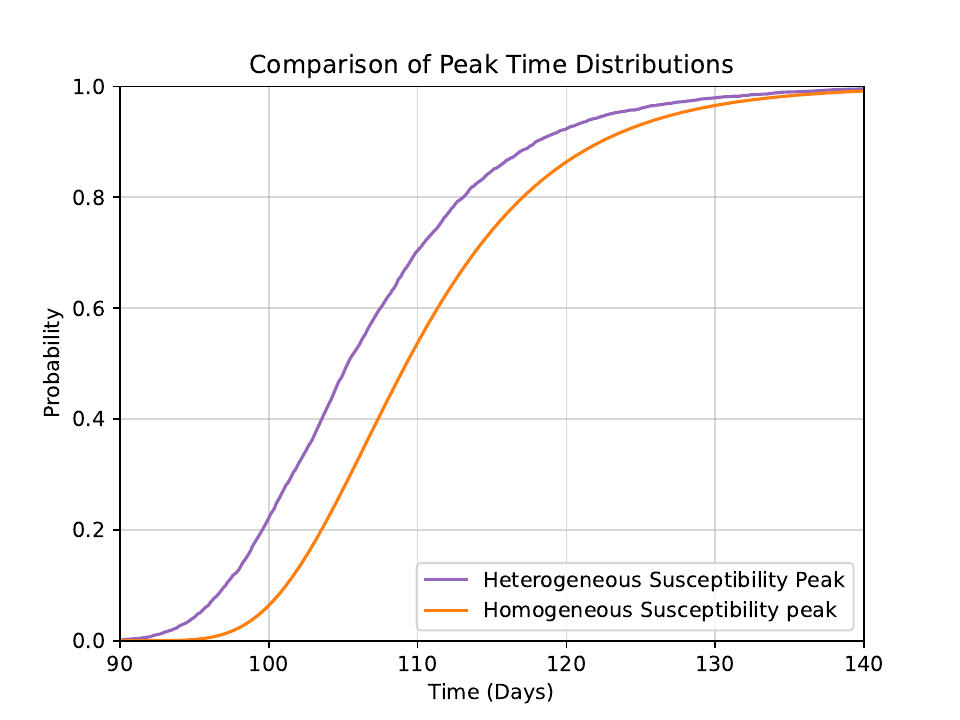}
    
    \caption{(Left column) True peak time distribution (estimated via 1000 simulations of the stochastic SIR model with homogeneous (top, left) and heterogeneous (bottom, left) susceptibility using the Gillespie algorithm) compared with approximated peak time distribution based on the approximating the early growth phase with using Feller's diffusion followed by a deterministic approximation once $Z^* = 125$ cases have been reached. (Right column) Comparison of dynamics for the SIR model with homogeneous and heterogeneous susceptibility profiles. Including heterogeneous susceptibility has a much larger effect on the size of the peak (shown above, with the peak indicated by the dashed lines) than on the distribution of the peak timing (shown below).}
    \label{fig:true_peak}
\end{figure}

To confirm the accuracy of our results in capturing the true underlying peak time distribution, we compare the approximated peak timing distribution with the distribution obtained by simulation of the full stochastic SIR epidemic. Details of the simulation and model are provided in Appendix \ref{truepeak_timing}. Plots of the approximate and simulated distributions are shown in Figure \ref{fig:true_peak}. 

\begin{figure}
    \centering
    \includegraphics[scale=0.6]{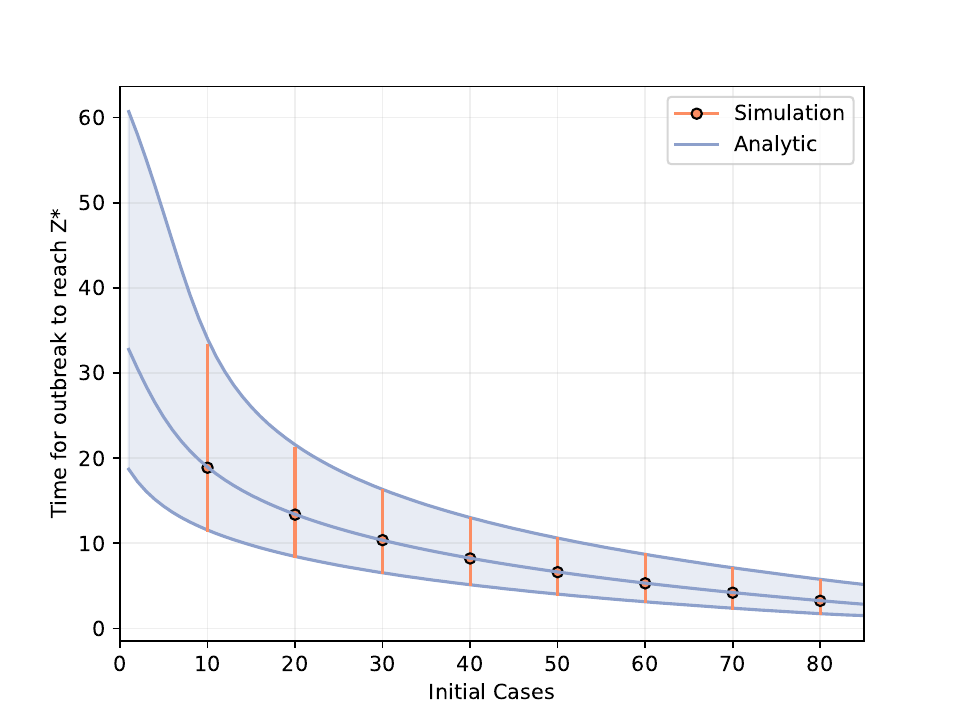}
    \caption{Uncertainty in the First Passage Time distribution to $Z^* = 125$ cases for different initial numbers of cases. The top and bottom (grey) lines correspond to analytic results for the 99\ts{th} and 1\ts{st} percentiles of the approximated distribution, respectively, with the middle line representing the mean of the distribution. Also shown are the First Passage Times to $Z^*$ obtained through simulation, with the range again shown between the  1\ts{st} and 99\ts{th} percentiles, and with the circle representing the mean. As the number of initial cases increases, both the mean time for the epidemic to reach $Z^*$ and the uncertainty around this mean decrease.}
    \label{fig:IC_sensitivity}
\end{figure}

We also show the impact that choosing different initial conditions has on the uncertainty in the FPT distribution. With a larger number of initial cases, the epidemic is already closer to becoming large, which results in lower population uncertainty and, therefore, greater confidence in the time taken for the number of cases to reach $Z^*$. As a result, the time taken for the dynamics to be well described by a determinisitic model is reduced when the initial number of cases is higher. We demonstrate this effect using both simulations and analytic results based on the Feller diffusion approximation in Figure \ref{fig:IC_sensitivity}.  

\subsection{Heterogeneous Susceptibility Model}

For the Novozhilov model, the early disease dynamics are described by exactly the same branching process as for the standard SIR model, and so the corresponding FPT distribution is the same for both models. To demonstrate this, we compare the model results both with and without heterogeneous susceptibility for $R_0 = 2$ and, in the case of the Novozhilov model, where individual susceptibility is drawn from a Gamma distribution with $r = 1$ and $\theta = 2$, i.e. from a $\Gamma(1, 1/2)$ distribution.  The convergence of our results on the peak time distribution to the true distribution obtained through simulations is shown in Figure \ref{fig:true_peak}. The main difference for the Novozhilov model is that the peak occurs earlier than for the standard SIR model. There is also additional variation in the timing of the peak that occurs in the model with heterogeneous susceptibility. 

The inclusion of heterogeneous susceptibility in the model has a much larger effect on the size of the peak than on the timing. A comparison of the dynamics of the model with and without heterogeneous susceptibility, as well as of the resulting peak time distributions, is shown in Figure \ref{fig:true_peak}. In the model with heterogeneous susceptibility, the peak occurs on average on day 106, compared with day 111 for the standard SIR model. However, the impact on the size of the peak is much greater, with a reduction of approximately $10^6$ in the number of infectious individuals at the peak when heterogeneous susceptibility is included in the model, compared to the standard SIR model. Whilst the impact on the timing of the peak is not substantial, our results that heterogeneous susceptibility can be included straightforwardly into our approach with little cost in terms of accuracy. Simulating the full stochastic SIR model with host heterogeneity, however, is highly expensive computationally, with 1,000 simulations requiring over 6 days to run.

\section{Discussion}
We have introduced two methods for approximating the temporal distribution for an epidemic whose early growth phase can be modelled using a branching process to reach a certain number of cases. We determine a suitable number of cases that should be reached in order for a deterministic model to be appropriate, based on analytic properties of the branching process. Once we obtain this time threshold, we are able to calculate the distribution in times taken for the process to reach this level, which we then translate forward in time to obtain a distribution of peak times for a deterministic approximation that starts with a stochastic growth phase. 

Our first method uses the solution of the widely-used Feller process to approximate the dynamics of the branching process, for which we obtain exact expressions using the Fokker-Planck equation. The second method makes an additional approximation to the square root of the Feller process, which allows for a Gaussian process solution from which we obtain an approximate FPT distribution. 

The advantage of using our methods is threefold: Firstly, we show in Figure \ref{fig:convergence} that, for reaching a large number of cases, our methods approximate the true FPT distribution better than the distribution obtained via the Gillespie algorithm when only $10^3$ trajectories are simulated. Secondly, our methods provide explicit expressions for the approximate distributions of both the number of cases and the hitting times, which gives greater mathematical insight than simulation alone. Finally, our approach is well suited to model calibration, particularly when estimates of the FPT distribution are needed for many different combinations of parameter choices. Even compared to only $10^3$ trajectories obtained via the Gillespie algorithm, our methods require much less computation time to approximate the FPT distribution (see Table \ref{tab:comp_time}). Whilst not as close to the true underlying FPT distribution as the non-central $\chi^2$, the Gaussian Process approximation offers an advantage over the non~=central $\chi^2$ in that it is straightforwardly applicable to a wider range of processes than the ones considered in this paper. In particular, the authors are currently working to apply these approximations to multi~=type branching processes, but we also envisage that our Gaussian Process approximation will be useful in time-inhomogeneous settings and for more general branching processes.

We have applied our results for the FPT distribution to calculate the peak time for an epidemic that evades extinction in the early growth phase. We have also shown that our results can be readily applied to extensions of the simple SIR model where, for example, heterogeneity in the susceptibility of hosts is included. These extensions do not significantly change the early dynamics of the epidemic, and so our results for the First Passage Time distribution are still valid in these cases. Furthermore, we anticipate that our results can be applied in a much broader context than simply in mathematical epidemiology; indeed, branching processes have been used to model the growth of cell populations, multi~=strain dynamics, phylogenetic trees as well as many other processes in biology and physics \cite{dyson2021possible, uecker2014evolutionary, hohna2016tess, arino2022bistability, kimmel2015branching}. 

Our methods presented in this paper enhance epidemic modelling by accounting for the uncertainty in the peak timing, but they can also help modellers to quantify the uncertainty due to parameter choices. Our results are obtained in only a fraction of the computation time taken to simulate the peak timing distribution using the full stochastic SIR model, which makes our methods more suitable for conducting grid searches of parameter space in order to quantify the parameter uncertainty in key model outcomes. We anticipate these approximations being used for scenario planning, where a number of different potential outcomes need to be considered in order to provide insights for operational planning. In particular, our results enhance deterministic models in this respect by providing a time window in which the peak number of infections is likely to occur. 
\section*{Code availability}

The code needed to reproduce the results shown here is available at \url{https://github.com/JCurran-Sebastian/FirstPassageTime_Branching}

\section*{Acknowledgements}

JCS acknowledges support from the Engineering and Physical Sciences Research Council (EPSRC) and from the Danish National Research Foundation (DNRF) via the Chair Grant awarded to Professor Samir Bhatt. LP was supported by the Wellcome Trust and the Royal Society (grant no. 202562/Z/16/Z). TH was supported by the Royal Society (grant no. INF/R2/180067). IH was supported by the National Institute for Health Research Policy Research Programme in Operational Research (OPERA, PR-R17-0916-21001) IH, LP and TH are supported by The Alan Turing Institute for Data Science and Artificial Intelligence, EPSRC (EP/V027468/1) and by UKRI through the JUNIPER modelling consortium (grant no. MR/V038613/1).

\section*{Conflicts of Interest}
The authors declare no conflicts of interest.


\bibliographystyle{IEEEtran}
\bibliography{main.bib}

\newpage

\appendix
\section*{Appendices}
\section{Probability of Extinction for the Single-Type Branching Process} \label{q_solve}
Here we solve the Chapman-Kolmogorov backward equation \eqref{Qeq}:
\begin{equation*}
    \frac{\partial Q}{\partial t} = \beta Q^2 -\rho Q + \gamma,
\end{equation*}
subject to $Q(0, s) = s$ in order to obtain the probability $q(t)$ that an outbreak that starts with an initial case at time $t=0$ has gone extinct by time $t$. We note that \eqref{Qeq} is a Riccati equation, which can be solved by substitution. We first note, that $Q(t, s) \equiv 1$ solves the ODE, and so the general solution takes the form $Q(t, s) = 1 + u(t, s)$, where $u(t, s)$ satisfies the first order ODE:
\begin{equation}
    u' - (\beta - \gamma)u = \beta u^2 .
\end{equation}
Making the substitution $u = 1/{z(t, s)}$ gives the linear ODE:
\begin{gather*}
    z' + (\beta - \gamma)z = -\beta \\ 
    \Rightarrow z(t, s) = \frac{\beta}{\gamma - \beta} + A{\mathrm{e}}^{(\gamma - \beta)t} .
\end{gather*}
Writing $Q(t, s) = 1 + (1/z)$ and using the initial condition $Q(0, s) = s$ to eliminate the constant of integration $A$, we obtain the following expression for the generating function:
\begin{equation}
    Q(t, s) = \frac{\gamma (s-1) - {\mathrm{e}}^{(\gamma - \beta)t}(\beta s - \gamma)}{\beta (s-1) - {\mathrm{e}}^{(\gamma - \beta)t}(\beta s - \gamma)}
\end{equation}
Finally, setting $s=0$ in the above expression yields the expression for $q(t)$ given in \eqref{Qsol}:
\begin{equation}
    q(t) = \frac{\gamma({\mathrm{e}}^{(\gamma - \beta)t} - 1)}{\gamma{\mathrm{e}}^{(\gamma - \beta)t} - \beta}
\end{equation}
Note that $q := \lim_{t \to \infty} q(t) = {\gamma}/{\beta}$ is the probability that an outbreak that begins with a single infectious individual ultimately goes extinct. 

\section{Solution to the Fokker-Planck equation for the Single-Type Branching Process} \label{Feller_solve}
We derive the solution of \eqref{FellerPDE} by first taking the Fourier transform:
\begin{equation*}
    \Tilde{f}(t, k) = \int_0^\infty f(t, x) {\mathrm{e}}^{-ikx} \, {\mathrm{d}}x
\end{equation*}
and then solving the resulting PDE via the method of characteristics. We first note the following properties of Fourier transforms:
\begin{gather}
    \widetilde{\left(\frac{\partial f}{\partial x}\right)} = ik \Tilde{f}(t, k) \\ 
    \widetilde{(xf(t, x))} = i \frac{\partial \Tilde{f}}{\partial k}
\end{gather}
and then, taking the Fourier transform of $f(t, x)$ in \eqref{FellerPDE}, we arrive at the equation:
\begin{gather}
    \frac{\partial \Tilde{f}}{\partial t} = \left(rk - \frac{i\rho k^2}{2} \right) \frac{\partial \Tilde{f}}{\partial k}, \label{FourierPDE} \\  \vspace{6pt}
    {\text{subject to:}} \quad \Tilde{f}(0, k) = {\mathrm{e}}^{-ik x_0}. \nonumber
\end{gather}
We now solve the above equation using the method of characteristics. Our aim is to find equations for the curves that lie in the surface $\Tilde{f}$ along which the value of $\Tilde{f}$ is constant. These are the characteristic curves of the PDE \eqref{FourierPDE}, parameterised by $s$, and are given by $(t(s), k(s))$ such that the tangent vector $\nabla(t(s), k(s))$ has coefficients that satisfy:
\begin{align}
    \frac{{\mathrm{d}}t}{{\mathrm{d}}s} & = 1 , & 
    t(s=0) & = 0 , \nonumber \\ 
    \frac{{\mathrm{d}}k}{{\mathrm{d}}s} & = -rk + \frac{i\rho k^2}{2} , & 
    k(s=0) &= k_0 \nonumber \\ 
    \frac{{\mathrm{d}}\Tilde{f}}{{\mathrm{d}}s} & = 0 , & 
    \Tilde{f}(t(0), k(0)) & = {\mathrm{e}}^{-i k_0 x_0}. 
    \label{characteristics}
    \end{align}
From the third equation, we see that our solution $\Tilde{f}(t, k)$ is constant along these characteristic curves. The first equation implies that $t = s$, whilst, for the second, we have that:
\begin{gather}
    \int_{k_0}^k \frac{{\mathrm{dv}}}{rv - \frac{i\rho v}{2}} = -\int_0^s {\mathrm{ds}} \nonumber \\ 
    \Rightarrow \frac{[\log(k) - \log(r - \frac{i\rho k}{2}) - \log(k_0) + \log(r - \frac{i\rho k_0}{2}) ]}{r} = -t. 
\end{gather}
Rearranging the above and isolating $k_0$, we have that:
\begin{equation}
    k_0 = \frac{k {\mathrm{e}}^{rt}}{1 + \frac{i \rho k}{2r}\left( {\mathrm{e}}^{rt} - 1\right) }.
\end{equation}
Finally, substituting our expression for $k_0$ into the final equation of \eqref{characteristics}, we obtain an expression for our solution $\Tilde{f}(t(s), k(s)) = \Tilde{f}(t(s=0), k(s=0))$:
\begin{equation*}
    \Tilde{f}(t, k) = {\mathrm{exp}} \left[ \frac{-ikx_0 {\mathrm{e}}^{rt}}{1 + \frac{i \rho k}{2r}\left( {\mathrm{e}}^{rt} - 1\right)} \right].
\end{equation*}
In order to simplify our expression for the Fourier transform of $f(t, x)$, we now scale $x$ so that $x \to \frac{4rx}{\rho \left( {\mathrm{e}}^{rt} - 1\right)}$. Finally, in order to obtain the characteristic function from the Fourier transform of $f(t, x)$, we also make the substitution $k \to -k$ so that we have:
\begin{gather}
    \Tilde{f}(t, k) = {\mathrm{exp}} \left[ \frac{i \lambda k}{1 - 2ik} \right] \label{chisq_char} \\ 
    {\text{where}} \quad \lambda = \frac{4rx_0 {\mathrm{e}}^{rt}}{\rho \left( {\mathrm{e}}^{rt} - 1\right)}. \nonumber
\end{gather}
Equation \eqref{chisq_char} is the characteristic function for a $\chi^2$ distribution with zero degrees of freedom and non-centrality parameter $\lambda$, first described by A. Siegel in \cite{siegel1979noncentral}, whose p.d.f. is given by:
\begin{equation}
    g(x; \lambda) = \frac{1}{2} \sqrt{\frac{\lambda}{x}} {\mathrm{e}}^{-\frac{1}{2}(\lambda + x)} I_1(\sqrt{\lambda x}),
\end{equation}
where $I_1(\cdot)$ is the modified Bessel function of the first kind.
The p.d.f. for for the number of cases at time $t$ in the Feller diffusion is therefore given by: 
\begin{equation*}
    f(t, x) = \frac{r}{\frac{\rho}{2}({\mathrm{e}}^{rt} - 1)} \sqrt{\frac{{\mathrm{e}}^{rt}}{x}} I_1 \left( \frac{2r\sqrt{x{\mathrm{e}}^{rt}}}{\frac{\rho}{2}({\mathrm{e}}^{rt} - 1)} \right) {\mathrm{exp}}\left(- \frac{r ({\mathrm{e}}^{rt}+x)}{\frac{\rho}{2}({\mathrm{e}}^{rt} - 1)}\right), 
\end{equation*}
where we have used the fact that, for a random variable $X$ with p.d.f. $f(x)$ and for a constant $c$ independent of $x$, the p.d.f. of $cX$ is given by $\frac{1}{c} \cdot f(\frac{x}{c})$.

\section{Peak Timing for the Stochastic SIR Epidemic} \label{truepeak_timing}
In order to verify the converegence of our results, we compare our results on the peak timing distribution for the branching process using the Gillespie algorithm followed by a deterministic approximation once the threshold $Z^* = 125$ cases has been reached, with those obtained by simulating the full stochastic SIR epidemic. For the stochastic SIR epidemic, we have the transitions:
\begin{gather}
    (S, I) \to (S-1, I+1) \quad {\text{with rate}} \quad \frac{\beta S I}{N} \nonumber \\ 
    (S, I) \to (S, I-1) \quad {\text{with rate}} \quad \gamma I. \nonumber
\end{gather}
As for the branching process, we simulate trajectories of the stochastic SIR epidemic using the Gillespie algorithm. Due to the large computation cost of performing these simulations with large numbers of cases, we run only 1,000 simulations each for the standard SIR model and the model with heterogeneous susceptibility. A comparison of the FPT distributions based on the branching process and the full stochastic SIR model is shown in the main text in Figure \ref{fig:true_peak}.

\end{document}